\documentclass[a4paper,12pt]{article}
\usepackage{psfrag}
\usepackage{mathrsfs} 

\usepackage{graphicx}
\usepackage{subfigure}
\usepackage[margin=10pt,font=small,labelfont=bf,labelsep=period]{caption}
\usepackage{amsmath}
\usepackage{amssymb}
\usepackage{dsfont} 
\usepackage[bottom]{footmisc} 
\usepackage{upref} 
\usepackage{cite} 
\numberwithin{equation}{section}

\renewcommand{\baselinestretch}{1.3}
\def\hgg{\widehat{g}}
\def\hgd{\widehat{g}_{\mu\nu}}
\def\hgp{\widehat{g}^{\mu\nu}}

\def\ed{\text{e}^{2\sigma}}
\def\eq{\text{e}^{4\sigma}}
\def\lp{\ell_{\rm Pl}}
\def\mp{m_{\rm Pl}}

\def\B{{\square}}
\def\S{{\cal S}} 
\def\16pig{16\pi G}
\def\8pig{8\pi G}
\def\32pig{32\pi G}
\def\96pig{96\pi G}

\def\be{\begin{equation}}
\def\ee{\end{equation}}  
\def\ba{\begin{eqnarray}}
\def\ea{\end{eqnarray}}  
\def\ie{{\it i.e.\,}}
\def\mp{m_{\rm Pl}}

\begin{document}
\begin{titlepage}
\renewcommand{\baselinestretch}{1.1}
\title{\begin{flushright}
\normalsize{MITP/13-006}
\bigskip
\vspace{1cm}
\end{flushright}
Modulated Ground State of Gravity Theories
with Stabilized Conformal Factor}
\date{}
\author{A. Bonanno$^{1,2}$ and M. Reuter$^3$\\
{\small $^1$INAF, Osservatorio Astrofisico di Catania, Via S.Sofia 78, 95123 Catania, Italy} \\[-0.2cm]
{\small $^2$INFN, Sezione di Catania, Via S.Sofia 72, 95123 Catania, Italy} \\[-0.2cm]
{\small $^3$ Institute of Physics, University of Mainz, Staudingerweg 7, D-55099 Mainz, Germany}}
\maketitle\thispagestyle{empty}

\begin{abstract} 
We discuss the stabilization of the conformal factor by higher derivative terms in a conformally reduced $R+R^2$ Euclidean gravity theory.
The flat spacetime is unstable towards the condensation of modes with nonzero momentum,
and they ``condense" in a modulated phase above a critical value of the coupling $\beta$ of the $R^2$
term. By employing  a combination of variational, numerical and lattice methods we show that 
in the semiclassical limit the corresponding functional integral is dominated  
by a single nonlinear plane wave of frequency $\approx 1/\sqrt{\beta} \lp$. 
We argue that the ground state of the theory is characterized  by a spontaneous breaking of
translational invariance at Planckian scales.
\end{abstract}
\end{titlepage}

\newpage
\section{Introduction}
It is not uncommon in euclidean field theories that spatially inhomogeneous, \ie non-constant
field configurations give rise to a lower value of the action functional than homogeneous ones.
As a result, at least at a semiclassical level, inhomogeneous configurations are likely 
to dominate the functional integral and to determine the quantum vacuum state, $|0 \rangle$. 
In  this situation the main properties of the true quantum state often can be understood by a saddle point
expansion of the quantum fluctuations (\ie, the integration variable of the functional integral) about
a specific set of field configurations; the latter have a position-dependent, translation invariance-breaking value 
of the field variable, and they are degenerate with respect to their value of the action.
In the quantum theory which is approximated in this way, this ``condensation'' 
of spatially inhomogeneous modes contributes to certain expectation values $\langle 0 | {\cal O} | 0 \rangle \not = 0$
which are sensitive to the nonvanishing kinetic energy of the configurations dominating the functional integral.
Here ${\cal O}$ is  a scalar operator constructed from the fundamental fields; for instance 
${\cal O}=\partial_\mu \phi \partial^\mu \phi$ in a scalar model, or ${\cal O} = {\rm tr} (F_{\mu\nu}F^{\mu\nu})$ in Yang-Mills theory.
Such contributions are sometimes referred to as ``kinetic condensates'' 
\cite{larewe}. They are to be distinguished from the more familiar translational invariant ``potential condensates" 
which underlie the conventional Higgs mechanism;
there the functional integral is dominated by nonzero but spatially constant scalar field configurations.

Let us denote the fundamental fields collectively by $\Phi$, the classical action by $S[\Phi]$, its stationary point(s) by $\Phi_0$, 
\ie $\frac{\delta S}{\delta \Phi}[\Phi_0]=0$, and its global minimum by $\Phi_{\rm min}$. Then the inverse propagator which governs
small fluctuations about a configuration $\Phi_0$ is given by the Hessian $S^{(2)}[\Phi]\equiv \delta S[\Phi]/\delta \Phi\delta \Phi$ 
evaluated at $\Phi=\Phi_0$. If the operator $S^{(2)}[\Phi_0]$ has negative eigenvalues then there exist directions in the field  space
along which the action can be lowered. This leads to a run away behavior of certain fluctuation modes whose excitation (condensation)
brings the field close to the global minimum of the action, $\Phi_{\rm min}$. 
Fluctuations about $\Phi_{\rm min}$ instead should all be stable, \ie the Hessian at the global minimum, $S^{(2)}[\Phi_{\rm min}]$, 
has only non-negative eigenvalues. 

When one uses stability criteria in order to judge whether a theory is physically acceptable, one may not confuse the Hessian
at some arbitrary solution of the field equation, $S^{(2)}[\Phi_0]$, with the one at the absolute minimum, 
$S^{(2)}[\Phi_{\rm min}]$. Only the latter must have a positive semidefinite spectrum. In the example where $\Phi$ is the Higgs field,
the homogeneous configurations $\Phi_0(x)=0$ and $\Phi_{\rm min}=v$, say, are both stationary points, corresponding to the ``false"
and the ``true" vacuum respectively. But clearly only the $S^{(2)}[\Phi_{\rm min}]$ is positive; the Hessian $S^{(2)}[\Phi_0]$ has a negative 
eigenvalue corresponding precisely to homogeneous fluctuations which tend to drive the field from $\Phi=0$ to $\Phi=v$.

In theories with instabilities of the kinetic type the situation is conceptually similar, albeit more complicated technically. Here the 
transition from a false to the true vacuum involves not just a shift of the field by the vacuum expectation value $v$, as 
in the Higgs case, but rather an expansion about a position dependent field configuration. 

A well-known example of a kinetic condensate is the gluon condensate in Quantum Chromo-Dynamics (QCD). There, the classical
action functional $\frac{1}{2}\int d^4 x \, {\rm tr} \, (F_{\mu \nu} F^{\mu \nu})$ is minimized by gauge field configurations with  $F_{\mu\nu}=0$,
but the effective action has its minimum at $F_{\mu\nu}\not =0$. An early attempt at finding its global minimum is the Savvidy vacuum 
\cite{sav}, the approximation of a covariantly constant color magnetic field. While its action is indeed lower than that of the naive vacuum
with $F_{\mu\nu}=0$, it turned out unstable in the  infrared (IR), and it has been argued that the true vacuum should be spatially 
inhomogeneous. The complexity of the QCD vacuum state is reflected by nonperturbative contributions to 
$\langle 0 | {\rm tr} \, (F_{\mu \nu} F^{\mu \nu}) |0\rangle$ and similar expectation values of more complicated gauge and Lorentz-invariant operators.

Typical examples from statistical physics are materials described by a Landau free energy functional with a Lifshitz point
\cite{lifs} and,  among the classes of magnetic materials, the antiferromagnetic ones. 
These latter display a microscopic order characterized by anti-parallel sublattices of spins, so that their
global  moments are exactly equal but opposite. Below the N\'{e}el temperature the susceptibility obeys the Curie-Weiss law for paramagnets but
with a negative exchange interaction. The antiferromagnetic order can then show up as  spatial inhomogeneity  \cite{kojima}.
Similarly, in a superconductor, the superconducting order parameter  displays a modulated phase  in the presence of a strong magnetic field, the 
Fulde-Ferrell-Larkin-Ovchinnikov (FFLO) phase \cite{fflo}. (The antiferromagnetic phase in scalar field theories has also been described in \cite{janos}.)

The model we are interested in here is euclidean quantum gravity \cite{eugravbook}.
Besides perturbative nonrenormalizability, the other property the Einstein-Hilbert action
\be
S_{\rm EH}[g_{\mu\nu}]=-\frac{1}{\16pig} \int d^4 x \sqrt{g} \, R
\ee
is notorious for is its unboundedness below. Indeed, if we make the conformal factor of the metric explicit
by setting $g_{\mu\nu}\equiv \, {\rm exp} (2\, \sigma) \, \hgd$ we obtain for the action
\be
S_{\rm EH}[g_{\mu\nu}]=-\frac{1}{\16pig} \int d^4 x \sqrt{\hgg}\; e^{2\sigma} \left( \widehat{R} +6\, \hgp \partial_\mu\sigma \partial_\nu\sigma \right)   
\ee
and its value can become arbitrarily negative when $\sigma(x)$ varies rapidly.

However, it is easy to modify the action in such a way that it becomes impossible to lower the action below all bounds by exciting the
conformal factor; hereby all the successfully tested predictions of classical general relativity are retained. The simplest way consists in 
adding a term proportional to the square of the Ricci scalar, with a positive coefficient $\beta>0$ :
\be\label{1p2}
S[g]=S_{\rm EH}[g]+\beta\,\int d^4 x \sqrt{g} \, R^2
\ee
The $R^2$ term involves 4 derivatives of metric. In an expansion about flat space, the modified action implies an inverse propagator for the 
conformal factor which, schematically, is of the form $\Omega(-\B)=\B+{\B \B}/(2 M^2)$ or, in momentum space, 
\be\label{1p3}
\Omega(p^2)=-p^2+\frac{(p^2)^2}{2M^2}
\ee
\begin{figure}[t]
\begin{center}
\includegraphics[width=9cm]{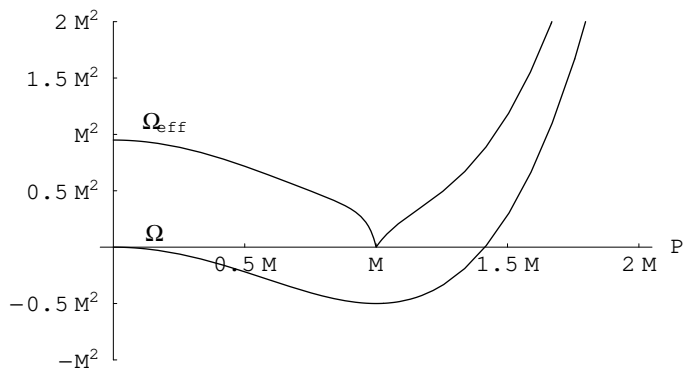}
\end{center}
\caption{The classical inverse propagator $\Omega(p^2)$ in eq.(\ref{1p3}) and the effective inverse propagator $\Omega_{\rm eff}(p^2)$ 
according to the toy model of eq.(\ref{1p10}). (Taken from \cite{larewe}.)}  
\label{flat}
\end{figure}
Here the $-p^2$ and $p^4$ terms are due to the Einstein-Hilbert and $R^2$ term, respectively, and $M\propto \mp/\beta^{1/2}$ is a constant of the
order of the Planck mass. For momenta much smaller than $M$ the $-p^2$ term dominates, and the field can lower its action by exciting such 
modes. If $p^2\gg M^2$, on the other hand, the positive $+p^4/ 2 M^2$ term dominates and it ``costs" action to excite the corresponding
modes. For euclidean momenta $p^2\equiv p_\mu p^\mu>0$ the kinetic operator $\Omega(p^2)$ is positive for 
$p^2> 2 M^2$; it has a minimum at $p^2=M^2$ where it assumes the value $\Omega( M^2)=-\frac{1}{2} M^2$. So, according 
to the modified action, the kinetic energy of the conformal factor is bounded below.

The important point is that the field configuration corresponding to flat space $(g_{\mu\nu} =\delta_{\mu\nu})$ cannot be at
the absolute minimum of $S$. In fact, flat space plays the role of a false vacuum here. While in absence of a cosmological constant it is
a stationary point of $S$, the Hessian $S^{(2)}[g_{\mu\nu}=\delta_{\mu\nu}]$ has negative eigenvalues which correspond precisely to the
fluctuation modes of the conformal factor with $\Omega(p^2)<0$. This is the typical symptom of a ``kinetic condensate" that wants to form
in order to lower the field's value of the action functional.
So in order to analyze both the classical and quantum properties of the theory based upon the modified action (\ref{1p2}) it is important
to have some understanding its minimum action configuration(s). 

\vspace{3mm}
Before we turn to this problem several remarks are in order.

\noindent
{\bf{(A)}} The perturbative quantization of $R^2$-gravity in an expansion about flat space results in a power counting renormalizable, 
though non-unitary theory \cite{stelle}. This well-known fact by no means rules out the existence of a bona fide quantum 
field theory based upon the $R^2$ action. It simply says that {\it flat space} is not the true ground state but only a false vacuum
of the theory; the negative-norm states (``ghosts") one encounters expanding about this false vacuum are a reflection of unstable
eigenmode of $S^{(2)}[g_{\mu\nu}=\delta_{\mu\nu}]$ which tend to grow when the system is heading for its true ground state.
The expansion about the true vacuum can very well be stable and unitary.

\noindent
{\bf (B)} The $R^2$- addition to the Einstein-Hilbert term is far from unique. In this paper we study  it as the simplest example 
of a theory in which the kinetic energy due to the conformal factor is bounded below.
The basic condensation mechanism should be similar in all models where $\Omega(p^2)=-p^2+f(p^2)$, with 
$f(p^2)>0$ a monotonically increasing function for $p^2\rightarrow \infty$. 

\noindent
{\bf (C)} Besides the conformal factor problem, the other key issue in trying to construct a quantum field theory of the metric is the 
ultraviolet (UV) renormalization. Even though this is not directly relevant for the investigations in this paper, 
we assume that the ultimate (stable) theory can be constructed along the lines of the Asymptotic Safety program \cite{wein}
based upon the gravitational average action \cite{mr}. The basic idea is to take the limit of an infinite UV cutoff at a non-Gaussian
renormalization group fixed point. 
The resulting field theory can then be predictive 
and well behaved on all scales. By now there is significant evidence for the existence of a suitable fixed point  
\cite{percadou,oliver1,frank1,oliver2,oliver3,oliver4,souma,frank2,prop,oliverbook,perper1,codello,litimgrav,frankmach,
BMS,oliverfrac,jan1,jan2,max,livrev,nagi,creh1,creh2,creh3,elisa,crehroberto,triat,saur2,dario}.
More than that, it has also been found \cite{oliver2} that the corresponding bare action seems to contain an $R^2$-term with a 
positive coefficient precisely what is needed for the envisaged condensation mechanism.

\noindent
{\bf (D)} In ref. \cite{larewe} the condensation of spatially inhomogeneous modes has been studied in detail within a scalar toy model
which mimics certain features of $R+R^2$ gravity. It consists of a complex scalar field $\chi$, on $d$-dimensional 
flat euclidean space, governed by the action
\be\label{1p10}
S[\chi]=\int d^d x \left \{ \chi^\ast \, \Omega (-\B) \, \chi +\frac{\lambda}{2} \, | \chi |^4 \right \}
\ee
The kinetic operator is the same as in Eq.(\ref{1p3}), so exciting the $\chi$-modes with momenta in the interval $p^2\in [0,2M^2]$
lowers the action. It was shown that the global minimum of the functional (\ref{1p10}) is given by the family of plane waves
\be\label{1p11}
\chi(x;n,\alpha)=\frac{M}{\sqrt{2\lambda}} \, \exp \left ( i M n_\mu x^\mu+i \alpha \right)
\ee
labeled by a unit vector $n\in S^{d-1}$ and a phase angle $\alpha$. The classical vacuum manifold is $S^{d-1}\times S^1$ therefore.

In ref.\cite{larewe} the dressed inverse propagator $\Omega_{\rm eff}(p^2)$ which appears in the model's effective action
functional $\Gamma[\chi]=\int \chi^\ast \Omega_{\rm eff}(-\B) \chi + \cdots$ was computed by a saddle point expansion about
the configurations (\ref{1p11}) which involved an integration over $n_\mu$ and $\alpha$. The result for 
$\Omega_{\rm eff}(p^2)$ is depicted in Fig.\ref{flat}. It shows quite nicely the {\it dynamical self-stabilization} of this theory by the condensation
of spatially inhomogeneous modes: due to the renormalization effects, the kinetic term has become positive semidefinite; for all modes
with $p^2\not=M^2$ it ``costs" energy (action, actually) to excite them. Only the modes with $p^2=M^2$ can be excited 
``for free"; this indicates that those modes might be unstable towards condensation.

The analysis of this toy model could be carried through in a rather complete way, including the calculation of its effective average action
$\Gamma_k[\chi]$ interpolating between $S$ and $\Gamma$. This was possible because of certain algebraic simplification arising
from the very special form of the classical action, and in particular since $\chi$ was taken to be a {\it complex} field.

The present paper is intended to be a first step towards a similar analysis for a realistic gravity action, with a real conformal factor in
particular. 

In the following we shall search for the global minimum of the $R+R^2$ action functional (with a cosmological constant included)
and try to establish its {\it modulated} nature for appropriate values of the parameters in the action. This is a very hard problem which cannot be solved exactly. 
In order to make progress, we restrict the domain of the functional to conformally flat metrics, \ie metrics conformal to
$\mathbb{R}^4$. We shall then use variational and numerical methods to find approximations to the global minimum in this subspace.

We believe that, as far as a possible dynamical resolution of the conformal factor problem is concerned, this 
restriction still contains the essential physics. In fact, the other (``transverse") metric degrees of freedom which we discard 
show no comparable instability and have no obvious reason to condense. 

We shall be particularly interested in finding {\it periodic} conformal factors which partition the spacetime into an array of elementary cells.
If they exist, they could possibly serve as a classical approximation of {\it quantum Minkowski space}, or its euclidean counterpart,
in the following sense: On microscopic, typically Planckian scales the metric is violently oscillating, but upon averaging it over a periodicity
volume it becomes perfectly flat, \ie after a purely classical ``coarse graining" one has $\langle g_{\mu\nu}(x) \rangle=\delta_{\mu\nu}$. 

The remaining sections of this paper are organized as follows. In Section 2 we discuss the general properties of the conformally reduced
$R+R^2$ action. In Section 3 we find its explicit global minimum for a special point in parameter space which saturates a kind
of Bogomolny bound. In Section 4 we employ variational techniques to obtain an approximation to the global minimum when 
the cosmological constant is zero, and in Section 5 we use numerical techniques to illustrate the impact which a nonzero cosmological constant
has on them.  Section 6 contains the conclusions.

\section{The conformal sector of $R+R^2$ gravity}
In the following we consider the euclidean action functional
\be\label{2p1}
\S[g_{\mu\nu}]=\int d^4 x \sqrt{g} \left \{ \frac{1}{\16pig}\left ( -R+2\Lambda\right) +\beta R^2\right \}
\ee
the cosmological constant $\Lambda$ and the dimensionless parameter $\beta$ are assumed positive throughout.
We shall see that $\beta>0$ guarantees  that $\S$ is bounded below. All stationary points of (\ref{2p1}), its global minimum in particular, satisfy
\be\label{2p2}
\frac{1}{\16pig}\left [ G_{\mu\nu}+\Lambda g_{\mu\nu} \right]+
\beta\left [ -(G_{\mu\nu}+R_{\mu\nu})R+2D_\mu D_\nu R -2g_{\mu\nu} D^2 R\right]=0
\ee
with the Einstein tensor $G_{\mu\nu}\equiv R_{\mu\nu}-\frac{1}{2} g_{\mu\nu} R$. Contracting (\ref{2p2}) with $g^{\mu\nu}$ yields
\be\label{2p3}
R-4\Lambda+(\96pig \beta) \; D^2 R=0
\ee

A special feature of the $R^2$ action (\ref{2p1}) is that {\it every stationary point of the Einstein-Hilbert action $(\beta=0)$ continues to be
a stationary point when the $\beta R^2$-term is added.}
In fact, it is easy to check that metrics satisfying the ordinary Einstein equations
\be\label{2p4}
G_{\mu\nu}=-\Lambda g_{\mu\nu}\;\;\;\;\;\; \Longleftrightarrow \;\;\;\;\;\; R_{\mu\nu}= \Lambda g_{\mu\nu}, \;\; R=4\Lambda
\ee
automatically solve  eq.(\ref{2p2}) also for $\beta\not=0$.

Mostly we shall be interested in those stationary points of ${\S}$ which are not already stationary points in absence of the stabilizing $R^2$ term.
As a consequence, (\ref{2p4}) will not be satisfied  in these cases, and eq.(\ref{2p3}) tells us that the curvature scalar is non-constant then.
More precisely, $R(x)$ may not be a harmonic function, $D^2 R\not = 0$.

By completing the square in eq.(\ref{2p1}) we can rewrite the action in the form
\be\label{2p10}
{\S}[g]=\int d^4 x \sqrt{g} \left \{ \beta \left ( R-\frac{4\Lambda}{1+\gamma} \right )^2 +\frac{\gamma}{(\32pig)^2 \beta} \right \}
\ee
with the abbreviation
\be
\gamma\equiv 128\pi\;  \beta G \Lambda -1 .
\ee
The representation (\ref{2p10}) is valid if $\beta G\Lambda\not = 0$, or $\gamma\not = -1$. Since, by assumption, $\beta>0$ it implies
a lower bound on the value of the action:\footnote{We consider compact manifolds without boundary or, in the infinite but periodic case, 
we refer the action to a single periodicity volume. Hence $\int d^4x \sqrt{g}$ is always finite.}
\be\label{2p11}
\S[g]\geq \frac{\gamma}{(\32pig)^2 \beta} \int d^4 x \sqrt{g}
\ee
Obviously the functional $\S$ is positive definite if $\gamma>0$, or
\be\label{2p12}
128\pi G \Lambda> 1 .
\ee

If $\gamma=0$, that is, for the special parameters values for which $128\pi \beta G \Lambda=1$, the minimum value of the action is at 
${\S}=0$, and the bound is saturated by metrics satisfying $R=4\Lambda$. While this is precisely the contracted form of the ordinary 
Einstein equations, here $R=4\Lambda$ has the interpretation of a kind of  ``Bogomolny equation". 

In the case $\gamma<0$ the functional ${\S}$ can assume negative values, but it is still bounded below.

\begin{figure}[t]
\centering
\includegraphics[width=10cm]{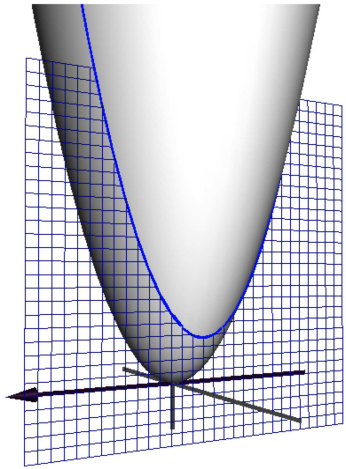}
\caption{Schematic view of the restricted functional space  studied in the paper. The paraboloid represents the action for 
the complete set of degrees of freedom, its global minimum at the origin thus satisfies the field equation for the $R+R^2$ model. 
The plane of frozen vector and tensor degrees of freedom cuts out instead a parabola (in bold in the figure) which represents the 
restricted functional space of the conformally reduced theory. (The axis with the arrow is the direction of the conformal field configurations.)
\label{cut}
}  
\end{figure}
Rather than trying to find the global minimum of ${\S}[g_{\mu\nu}]$ defined over the space of {\it all}
metrics we shall be more modest here and only analyze the action restricted to the conformally flat metrics
\be\label{2p15}
g_{\mu\nu}=\phi^2(x) \delta_{\mu\nu} .
\ee
Writing $S[\phi]=\S[g_{\mu\nu}=\phi^2 \delta_{\mu\nu}]$ for the functional depending on the conformal factor $\phi$
we obtain from (\ref{2p1}):
\be\label{2p16}
S[\phi]=\int d^4 x \left \{ \frac{3}{\8pig} \, \phi \B \phi+\frac{\Lambda}{\8pig} \, \phi^4+36\beta\left( \frac{\B\phi}{\phi}\right)^2\right\}
\ee
where we used that $\sqrt{g}=\phi^4$ and $R=-6\phi^{-3}\B \phi$ with  $\B=\delta^{\mu\nu}\partial_\mu\partial_\nu$
for metrics of the form $g_{\mu\nu}=\phi^2 \delta_{\mu\nu}$. 
If $\beta=0$ the restricted functional (\ref{2p16}) has the appearance of a scalar $\phi^4$-action with a ``wrong sign" kinetic term.

The main topic of he present paper is the investigation of the global minimum action configuration(s) $\phi_{\rm min}$ 
of the restricted functional $S[\phi]$, Eq.(\ref{2p16}). It is plausible to assume that its essential qualitative features, in particular
the existence of a modulated phase for certain parameter values, will be shared by the true minimum $g_{\mu\nu}^{\rm min}$,
\ie that of ${\S}[g_{\mu\nu}]$ defined for all, not necessarily conformally flat metrics. 

While we hope that $g_{\mu\nu}^{\rm min}$ is to some extent similar to the conformally flat metric with the lowest action, 
\be\label{2p17}
g_{\mu\nu}^{\rm conf-min}\equiv \phi^2_{\rm min}(x) \delta_{\mu\nu}
\ee
we emphasize that {\it $g_{\mu\nu}^{\rm conf-min}$ is not a stationary point of ${\S}[g_{\mu\nu}]$ in general.}
The metric (\ref{2p17}) is found from 
\be\label{2p18}
\frac{\delta}{\delta\phi(x)} S[\phi] \, {\bigg |}_{\phi = \phi^{\rm min}} \equiv 
\frac{\delta}{\delta\phi(x)}\S[g_{\mu\nu}=\phi^2 \delta_{\mu\nu}]\,{\bigg |}_{\phi=\phi_{\rm min}}
\ee
while $g_{\mu\nu}^{\rm min}$ satisfies (\ref{2p2}), \ie
\be\label{2.19}
\frac{\delta}{\delta g_{\mu\nu}(x)}\,\S[g]\,{\bigg |}_{g = g^{\rm min}}=0
\ee 
Fig.(\ref{cut}) is a graphical illustration of the fact that (\ref{2p17}) with (\ref{2p18}) does not imply  (\ref{2.19}) in general.
In fact, the metric $g_{\mu\nu}^{\rm conf-min}$ sits to the minimum of the parabola cut out by the plane, but it is not a stationary point of the 
unrestricted functional represented by the paraboloid.

\noindent
{\bf{(A)}} Before turning to the actual minimization problem a remark concerning dimensions might be in order. Throughout this paper,
the coordinates denoted $x^\mu$ are assumed dimensionless. Hence, in this system of coordinates, all metric components 
$g_{\mu\nu}$ have mass dimension $-2$, for $ds^2=g_{\mu\nu}dx^\mu dx^\nu$ has $[ds^2]=-2$ always.
As a result, the conformal factor introduced as in (\ref{2p15}) has the dimension of a length, $[\phi(x)]=-1$.

\noindent
{\bf{(B)}} Sometimes it is advantageous to introduce the conformal factor in a slightly different way so that the second order kinetic term in $S$
has a standard normalization (up to its sign). Writing
\be\label{2p21}
g_{\mu\nu}(x) = \frac{1}{3} (4\pi G) \, \chi^2(x) \, \delta_{\mu\nu}, \; \text{\ie} \; \phi^2\equiv \frac{1}{3}(4\pi G) \, \chi^2
\ee
the action for $\chi$ reads
\begin{eqnarray}
S[\chi] & = & \int d^4 x\Big \{ \frac{1}{2} \, \chi \B \chi + 36 \beta \, \Big ( \frac{\B \chi}{\chi}\Big )^2 + \frac{u}{4!} \, \chi^4 \Big \}  \label{2p22a}\\
            & = & 36\beta \int d^4 x \, \chi^4 \Big ( \frac{\B \chi}{\chi^3} +\frac{1}{144\beta}  \Big )^2+\frac{\gamma}{576 \beta}\int d^4 x \, \chi^4 \label{2p22}
\end{eqnarray}
with the parameters 
\be\label{2p23}
u\equiv \frac{16\pi}{3} G \Lambda,\,\,\,\,\,  \gamma=128\pi\;\beta G\Lambda-1\equiv 24 \; u \beta -1 
\ee
As $[G]=-2$, $[\Lambda]=+2$ the parameter $u$ is dimensionless, the same is true for the new field variables: $[\chi]=0$.
Since $\lp\equiv \sqrt{G}\equiv \mp^{-1}$ is the Planck length, we see that $\chi$ measures proper distances 
$ds^2=\frac{4\pi}{3} (\chi(x) \lp)^2 \, \delta_{\mu\nu} dx^\mu dx^\nu$ in units of $\lp$. 

\noindent
{\bf{(C)}} The above convention of dimensionless coordinates is most convenient when dealing with curvilinear coordinates, in curved
space in particular. It is, however, not the convention usually adopted in ordinary quantum field theory on Minkowski space.
There one prefers using cartesian coordinates with respect to which the metric components are normalized to $\pm 1$. Since 
$ds^2$ is still required to have dimension $-2$, those coordinates necessarily  have the dimension of a length.

The transition from our above conventions to this kind of dimensionful coordinates, henceforth denoted $\bar{x}^\mu$, is achieved by the rescaling 
\be\label{2resc}
\bar{x}^\mu \equiv c \, \sqrt{\frac{4\pi}{3}} \, \lp \, x^\mu, \;\;\;\;\;\;\;\;\; [\bar{x}^\mu]=-1 
\ee
In terms of the $\bar{x}$'s the line element implied by a function $\chi(x^\alpha)$ reads
\be\label{2line}
ds^2=\frac{1}{c^2} \, \chi^2 \Big ( x^\alpha = \sqrt{\frac{3}{4\pi}} \frac{\bar{x}^\alpha}{c \, \lp}\Big )\, \delta_{\mu\nu}d\bar{x}^\mu d\bar{x}^\nu
\ee 
Here $c>0$ is an arbitrary dimensionless constant which we may use to change the absolute normalization of the $\bar{x}$'s.  For instance, if 
$\chi^2$ happens to be a periodic function we might choose $c^2\equiv \langle \chi^2(\cdot)\rangle$ to be the average value of $\chi^2$ taken over one periodicity volume
(elementary cell). As a result, the line element if averaged correspondingly is precisely that of flat euclidean space expressed in terms of standard cartesian coordinates:
\be\label{2averaged}
\langle ds^2 \rangle = \delta_{\mu\nu} d\bar{x}^\mu d\bar{x}^\nu 
\ee
Later on we shall indeed find minimum action configurations $\chi(x)$ which have a Planck scale periodicity and average to flat space on macroscopic scales.

\noindent
{\bf{(D)}} Varying the action $S[\chi]$, eq.(\ref{2p22a}), with respect to $\chi$ yields the following 4th-order partial differential equation for its stationary points:
\be
\B \chi +\frac{u}{6}\, \chi^3 +72\beta\,\Big [\B \Big ( \frac{\B \chi}{\chi^2} \Big ) - \frac{(\B \chi )^2}{\chi^3} \Big ] = 0\label{2p24}
\ee
Besides the global minimum of $S[\chi]$ which we are after, this equation is satisfied also by many local minima and maxima and stationary points of mixed type. 
In particular all solutions $\chi_0$ of the equation (\ref{2p24}) with $\beta=0$ are solutions of the full equation with $\beta\not =0$, too. In fact, we see immediately that
\be\label{2p25}
\Box \chi_0+\frac{u}{6} \, \chi_0^3 = 0
\ee
implies $\beta [ \B (\chi_0^{-2}\B \chi_0) -\chi_0^{-3}(\B \chi_0)^2]=0$. Thus the solutions of this somewhat trivial type are insensitive to the value of $\beta$.

\noindent
{\bf{(E)}} In general the partial differential equation (\ref{2p24}), equipped with appropriate boundary conditions, is hard to solve. In Section 5 we shall 
analyze it using numerical methods. For that purpose, still another parametrization of the conformal factor, namely in terms of an exponential
turned out advantageous:
\be\label{2p26}
g_{\mu\nu}= \frac{1}{3} (4\pi G) \; \text{e}^{2\sigma(x)} \, \delta_{\mu\nu}, \; \; \text{\ie}\; \; \; \chi(x) = \text{e}^{\sigma(x)} .
\ee
The corresponding action reads
\ba\label{2p27}
&&S[\sigma]= \int d^4 x\, \Big \{ \frac{1}{2} \ed \big ( \B \sigma +\partial_\mu \sigma \partial^\mu\sigma \big )+36 \beta \big (\B \sigma +\partial_\mu \sigma \partial^\mu\sigma \big )^2+\frac{u}{4!} \, \eq \Big \}
\ea
and the condition for stationarity assumes the form 
\ba\label{2p28}
&&\ed \big [ \B \sigma  +\partial_\mu \sigma \partial^\mu\sigma \big ]+\frac{u}{6}\, \eq +72\beta \big [ \B\B\sigma+2(\partial_\mu\partial_\nu\sigma)(\partial^\mu\partial^\nu\sigma) -2(\B \sigma)(\B\sigma)\nonumber\\[2mm]
&&-2(\partial_\mu\sigma)(\partial^\mu\sigma)\B \sigma-4(\partial_\mu\sigma)(\partial_\nu\sigma)(\partial^\mu\partial^\nu\sigma)\big ]=0.
\ea

In the rest of this paper we shall approach the minimization problem of $S$ by three different methods: In Section 3 we study the special case $\gamma=0$ by means 
of the ``Bogomolny equation", in Section 4 we use a variational method, and in Section 5 numerical techniques. 
\section{Special case $\gamma=0$: The Bogomolny equation}
In this section we assume that the parameter combination $\gamma = 128\pi\beta G\Lambda -1$ assumes the special value $\gamma=0$. 
Then Eq.(\ref{2p10}) and Eq.(\ref{2p11}) imply that $\S[g_{\mu\nu}]\geq 0$ and equality holds for metrics with 
\be\label{3p1}
R(g) = 4\Lambda
\ee
We shall refer to (\ref{3p1}) as the ``Bogomolny equation". It happens to coincide with the contraction of the ordinary (\ie second order) Einstein equation.
If $\gamma=0$, all solutions of (\ref{3p1}) saturate the lower bound $\S[g_{\mu\nu}]=0$.

Here we shall consider only conformally flat solutions to Eq.(\ref{3p1}). Writing the metric as in Eq.(\ref{2p21}) and inserting it into (\ref{3p1}) we are thus led
to solve the Yamabe problem
\be\label{3p2}
\B \chi + \frac{u}{6}\, \chi^3 = 0
\ee
This equation has the same mathematical structure as Eq.(\ref{2p25}), its interpretation is somewhat different though. To get (\ref{2p25}) from (\ref{2p24})
we had set $\beta=0$, but (\ref{3p2}) refers to the specific nonzero $\beta$ for which $128\pi\,\beta G\Lambda= 1$. (Recall also that $u\equiv \frac{16\pi}{3} G \Lambda$
whence $u/6=1/(144\beta)$ if $\gamma=0$. Therefore Eq.(\ref{3p2}) coincides with the Bogomolny equation one reads off from the reduced action (\ref{2p22a}).)

The only known periodic solutions to equation (\ref{3p2}) are the euclidean analogs of traveling plane waves. If $\chi$ depends only on one of the coordinates, 
$x^1\equiv x$, say, Eq.(\ref{3p2}) can be interpreted as the Newtonian equation of motion of a particle moving along the ``$\chi$-axis" under the influence of a potential 
$V(\chi)=\frac{u}{24}\chi^4$ :
\be\label{3p3}
\chi'' (x) = -\frac{d}{d\chi}\, V(\chi)
\ee
Here the prime denotes a derivative with respect to $x$ which plays the role of time. Since the cosmological constant is assumed positive, $u$ and $V$
are positive too, so that the solutions to (\ref{3p3}) correspond to anharmonic oscillations. 
They can be found explicitly in terms of Jacobi elliptic functions \cite{abram}.

The general plane wave type solution of (\ref{3p2}) involves two free constants, $\widehat{\chi}$ and $\alpha$. It reads
\be\label{3p4}
\chi(x) = \widehat{\chi} \, \text{sn}\Big ( \widehat{\chi} \sqrt{\frac{u}{12}} \, n_\mu x^\mu + \alpha ; i \Big )
\ee
Here {\it sn} denotes the {\it sinus amplitudinis} with purely imaginary modulus $k=i$, and $n^\mu$ is an arbitrary unit vector, $\delta_{\mu\nu} n^\mu n^\nu=1$.
Obviously the solutions (\ref{3p4}) have a nontrivial periodicity in the direction of $n_\mu$ and are constant in the three directions perpendicular
to it. The coordinate length of the period in $n$-direction is \cite{grad}
\be\label{3p5}
\Delta x = \frac{8 \sqrt{3} K(i)}{\widehat{\chi} \sqrt{u}}
\ee
with ${\rm K}(i)$ an elliptic integral of the first kind: 
\be\label{3p6}
{\rm K}(i) = \int_0^1 \frac{dt}{\sqrt{1-t^4}} = \frac{1}{4\sqrt{2\pi}} \Big [ \Gamma (\frac{1}{4}) \Big ]^2  \approx 1.31
\ee

If we employ the dimensionful coordinates 
\be\label{3p7}
\bar{x}^\mu = \zeta \,  \widehat{\chi} \sqrt{\frac{4\pi}{3}} \lp \, x^\mu
\ee
the line element related to the minimum action configuration (\ref{3p4}) reads
\be\label{3p8}
ds^2 = \frac{1}{\zeta^2} \; \text{sn}^2 \Big ( \sqrt{\frac{u}{\pi}} \, \frac{n_\mu \bar{x}^\mu}{4\zeta\lp}+\alpha; i \Big ) \, \delta_{\mu\nu} d \bar{x}^\mu d \bar{x}^\nu
\ee 
Here $\zeta^2\equiv \langle \text{sn}^2(\cdot; i)\rangle $ is taken to be the average, over one period, of the squared Jacobi function:
\be\label{3p9}
\zeta^2 = {\rm E}(i)/{\rm K}(i) -1 \approx 0.45
\ee
where ${\rm E}(i)$ is an elliptic integral of the second kind, 
\be\label{3p62}
{\rm E}(i) = \int_0^1 {dt} \sqrt{\frac{1+t^2}{1-t^2}} \approx 1.91
\ee

With this particular normalization of the coordinates, {\it the average of $ds^2$ equals the standard form of the line element on flat euclidean space, 
\be\label{3p10}
\langle ds^2 \rangle = \delta_{\mu\nu} d \bar{x}^\mu d \bar{x}^\nu,
\ee
even though the scale factor is rapidly oscillating on short scales.} 
If $u=O(1)$, the coordinate length $\Delta \bar{x}$ of one period is of the order of a Planck length. 

\begin{figure}[h]
\centering
\includegraphics[width=8cm]{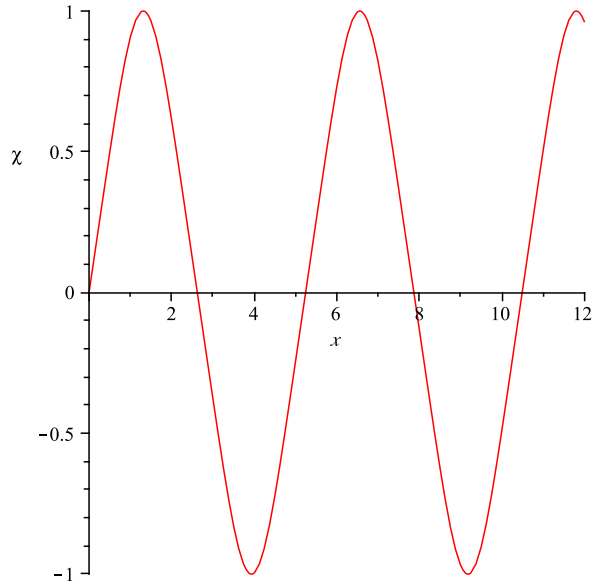}
\caption{The solution  (\ref{3p4}) for $\widehat{\chi}=1$, $u=12$ and $n_\mu = \delta_{\mu 1}$. \label{jb}}  
\end{figure}

The conformal factor pertaining to Eq.(\ref{3p8}) is depicted in Fig.(\ref{jb}).
A potentially unphysical feature of (\ref{3p8}) is that the metric degenerates at the zeros of the Jacobi function.

\section{The variational approach}
In this section we employ a variational technique in order to minimize the restricted action functional (\ref{2p22a}). Here we shall focus on the case 
of vanishing cosmological constant, $\Lambda=0$, whence $u=0$, and
\be
S[\chi] = \frac{1}{2} \int d^4 x \Big \{  -(\partial_\mu \chi)^2 + \frac{1}{Q^2}\Big ( \frac{\B \chi}{\chi}\Big )^2 \Big \} \label{4p1}
\ee
where we abbreviated
\be
Q\equiv( 72\beta)^{-1/2}\label{4p2}
\ee
The impact of the cosmological constant will be explored later on with a different method.
\subsection{The single plane-wave ansatz}
We start by using a variational ansatz in which the modulation of the conformal factor has the structure of a single plane wave. Without loss of 
generality we may use a frame such that the corresponding wave vector points along the $x^3\equiv z$ direction. Thus the trial ansatz for $\chi$ depends on a single
cartesian coordinate only:
\be
\chi(z)=A \Big [1+h \cos(\nu z)\Big ]\label{4p3}
\ee
Eq.(\ref{4p3}) comprises a 3 parameter family of conformal factors. They are labeled by the variational parameters $A$, $h$ and $\nu$ which we are going to adjust
in such a way that (\ref{4p1}) restricted to the trial space assumes its minimum.

By inserting (\ref{4p3}) into (\ref{4p1}),  we find 
\be
S= {\rm Vol} \cdot \mathscr{S}  \;\; {\rm with} \;\;\; \mathscr{S}=  \frac{\nu}{2\pi} \int_0^{2\pi/\nu} dz
\Big \{  -(\partial_\mu \chi)^2 + \frac{1}{Q^2}\Big ( \frac{\B \chi}{\chi}\Big )^2 \Big \}
\ee
where Vol is a (very large) four-dimensional normalization volume element, and  $\mathscr{S}$ denotes the action, per transverse 3-volume, 
averaged over one period $2\pi/\nu$ in the $z$-direction. Explicitly, the integral for  $\mathscr{S}$ simplifies to 
\ba
&& \mathscr{S} =\frac{\nu}{2\pi} \int_0^{2\pi/\nu} dz \frac{h^2\nu^2}{2{Q}^{2} \left( 1+2\,h\,\cos \left( {\nu}\,z \right) +{h}^{2}\cos \left( {\nu}\,z \right)^{2} \right)} 
\label{zir0}\\[2mm]
&&\times \Big [-{A}^{2}{Q}^{2}+{A}^{2}{Q}^{2} \cos \left( {\nu}\,z \right)^{2}-2\,{A}^{2}{Q}^{2}h\,\cos \left( {\nu}\,z \right) +2\,{A}^{2}{Q}^{2}h\,  \cos \left( {\nu}\,z \right)^{3}\nonumber\\[2mm]
&&-{A}^{2}{Q}^{2}{h}^{2}  \cos \left( {\nu}\,z \right)^{2}+{A}^{2}{Q}^{2}{h}^{2} \cos \left( {\nu}\,z\right)^{4}+\cos \left( {\nu}\,z \right)^{2}{{\nu}}^{2}\Big]\nonumber
\ea
Performing the integrations yields the exact result
\ba
&& \mathscr{S} =\frac{\nu^2}{4 Q^2 (1-h^2)^{3/2}} \\
&&\times\Big [ h^4 A^2 Q^{2}\sqrt {1-h^2}+4\nu^2 h^2-h^2 A^2 Q^2\sqrt {1-{h}^{2}}-2\,{{\nu}}^{2}\sqrt {1-{h}^{2}}{h}^{2}-2{{\nu}}^{2}+2{{\nu}}^{2}\sqrt {1-{h}^{2}} \Big ] \nonumber
\ea
From this expression we obtain the following partial derivatives with respect to the two variational parameters $h$ and $\nu$:
\ba
&&\frac{\partial  \mathscr{S}}{\partial h} = 
-\frac{h \nu^2} {2 Q^2 (1-h^2)^{5/2}}\label{hh}\\[2mm]
&&\times \Big [-2\,{h}^{2}{A}^{2}{Q}^{2}\sqrt {1-{h}^{2}}+{h}^{4}{A}^{2}{Q}^{2}\sqrt 
{1-{h}^{2}}+{A}^{2}{Q}^{2}\sqrt {1-{h}^{2}}-{{\nu}}^{2}-2\,{h}^{2}{
{\nu}}^{2}\Big ],\nonumber\\[2mm]
&&\frac{\partial  \mathscr{S}}{\partial \nu}  = 
-\frac{\nu} {2 Q^2 (1-h^2)^{3/2}}\label{hnu}\\[2mm]
&&\times
\Big [
{h}^{4}{A}^{2}{Q}^{2}\sqrt {1-{h}^{2}}-{h}^{2}{A}^{2}{Q}^{2}\sqrt {1-{
h}^{2}}+8\,{h}^{2}{{\nu}}^{2}-4\,{{\nu}}^{2}\sqrt {1-{h}^{2}}{h}
^{2}-4\,{{\nu}}^{2}+4\,{{\nu}}^{2}\sqrt {1-{h}^{2}} \Big].\nonumber
\ea
The minima of the action can now be searched by equating (\ref{hh}) and (\ref{hnu}) to zero and solving 
for $h$ and $\nu$. After some manipulation,  the following solution is obtained:
\be
h=0.5830, \;\;\;\;\;\;\; \nu = 0.4595 \, A\, Q,  \,\,\,\,\,\,\,\, S[\chi]/{\rm Vol}=-0.008960 \, A^4 Q^2 \label{s1} .
\ee

This solution amounts to the non-degenerate, global minimum of ${\mathscr S}(h,\nu, A)$ considered a function of $h$ and $\nu$ only, with
$A$ kept fixed. Indeed, the variational principle does not fix the overall normalization $A$; as we shall see in a moment it will assume
a unique value once we adopt standard coordinates on the flat euclidean space which arises in the macroscopic limit.

Before closing this subsection we comment on a technical issue.
Due to the complicated denominator of the integrand in (\ref{zir0}) it is difficult to extend this approach to more complicated test functions
since usually the integral cannot be performed exactly any longer. Let us therefore use the exactly soluble case (\ref{zir0})
in order to test an alternative strategy, namely  to expand the integrand in (\ref{zir0}) in a power series of $h$ 
up to a given finite order prior to the $z$-integration. This latter integration is then performed analytically on a polynomial expression of trigonometric functions.
The expansion is found to rapidly  converge to the values (\ref{s1}) when we increase the order of the polynomial,
as it is possible to notice in Tab.(\ref{tab:hresult}).

\begin{table}[h]
\caption{Convergence} 
\centering 
\begin{tabular}{cccc} 
\hline           
order  & $h$ & $\nu/ AQ$ &  $\mathscr{S}/A^4Q^2$ \\ 
\hline
$O(h^{8})$ & 0.7520 & 0.3908 &  -0.009884 \\
$O(h^{10})$ & 0.6337 & 0.4339 &  -0.009317 \\
$O(h^{12})$ & 0.6016 & 0.4489 &  -0.009094 \\
$O(h^{14})$ & 0.5902 & 0.4550 &  -0.009010 \\
$O(h^{16})$ & 0.5857 & 0.4579 &  -0.008978 \\
\hline                         
\end{tabular}
\label{tab:hresult}
\end{table}

\subsection{The combination of two plane-waves}
The important question is to understand if a combination of several plane-waves can lower further the value of the action. In order to address this point 
a more general trial ground state needs to be considered. In particular, if we set
\be
\chi(x)=A \Big [1+h \Big ( r  \cos(\omega z) +\cos(\nu z + s y) \Big) \Big ] 
\ee
we can discuss the possibility that a combination of two plane-waves attains a lower value for the action by considering 
simultaneous variations of $h$, $r$, $\nu$, $\omega$ and $s$. Without loss of generality, the wave vector of the
first plane wave has a $z$-component only, and that of the second lies within the $z-y$ plane.

In this case it is more convenient to perform the $z$ integration after the expansion in powers of $h$ of the integrand, as suggested before. The action density thus becomes  
\ba
&&S[\chi]/{\rm Vol} =\frac{\Delta_0+h^2(\Delta_2 s^2 +\Delta_4 s^4)}{32 Q^2}
\label{vor}
\ea
where we find for the three coefficients, to order $O(h^6)$,  
\ba
&&\Delta_0=8\,{\nu}^{4}+18\,{h}^{2}{\nu}^{4}+25\,{h}^{4}{\nu}^{4}+90\,{h}^{4}
{\nu}^{4}{r}^{2}+15\,{h}^{4}{\nu}^{4}{r}^{4}+25\,{h}^{4}{r}^{6}{
\omega}^{4}+15\,{h}^{4}{r}^{2}{\omega}^{4}\nonumber\\
&&+12\,{h}^{2}{\nu}^{4}{r}
^{2}+18\,{h}^{2}{r}^{4}{\omega}^{4}-8\,{Q}^{2}{A}^{2}{\nu}^{2}-8\,
{Q}^{2}{A}^{2}{r}^{2}{\omega}^{2}+12\,{h}^{2}{r}^{2}{\omega}^{4}+
8\,{r}^{2}{\omega}^{4}\nonumber\\
&&+120\,{h}^{4}{r}^{2}{\omega}^{2}{\nu}^{2}+120
\,{h}^{4}{r}^{4}{\omega}^{2}{\nu}^{2}+90\,{h}^{4}{r}^{4}{\omega}^{
4}+48\,{h}^{2}{r}^{2}{\omega}^{2}{\nu}^{2}\\[4mm]
&&\Delta_2=
120\,{h}^{4}{r}^{4}{\omega}^{2}{\nu}^{2}+36\,{h}^{2}{\nu}^{4}+180
\,{h}^{4}{\nu}^{4}{r}^{2}+16\,{\nu}^{4}+30\,{h}^{4}{\nu}^{4}{r}^{4}+120\,{h}^{4}{r}^{2}{\omega}^{2}{\nu}^{2}\nonumber\\
&&+50\,{h}^{4}{\nu}^{4}+24
\,{h}^{2}{\nu}^{4}{r}^{2}-8\,{Q}^{2}{A}^{2}{\nu}^{2}+48\,{h}^{2}{
r}^{2}{\omega}^{2}{\nu}^{2}\\[4mm]
&&\Delta_4=
90\,{h}^{4}{\nu}^{4}{r}^{2}+8\,{\nu}^{4}+18\,{h}^{2}{\nu}^{4}+15\,
{h}^{4}{\nu}^{4}{r}^{4}+25\,{h}^{4}{\nu}^{4}+12\,{h}^{2}{\nu}^{4}{r}^{2}
\ea

The minimization procedure involves a solution of a rather involved system of five nonlinear equations. It turns out that 
the only real solutions are  the previous single plane wave solutions 
\be
h=0.7521,\;\;\;\;\; \nu = 0.3908\, AQ, \;\;\;\;\;\;\;\, \omega={\rm arbitrary}, \;\;\;\;\;\;\; r=0, \;\;\;\; s=0 \label{sp1}
\ee
and a class of coupled plane-waves labeled by the parameter $s$,
\be
h=0.4229,\;\;\; \nu = \sqrt{\frac{500-135\sqrt{5}}{1271(1+s^2)}}\, AQ, \;\;\omega=0.3948\,AQ, \;\;\;  r=1, \; s={\rm arbitrary} \label{sp2}
\ee
However, we find that the value of the action is lower (more negative) for the single plane wave solution (\ref{sp1})  than for 
for multi plane-wave solution (\ref{sp2}), being $-0.0108 A^4 Q^2$ in the first case and $-0.00697 A^4 Q^2$ in the second case.
We have checked that this conclusion is further reinforced by the inclusion of additional powers of $h$ in the determination of the minimum up to $O(h^{12})$.

\subsection{The optimal trial metric}
Summarizing the above results for the case $\Lambda=0$ we can say that the variational calculations indicates that the absolute
minimum is a {\it single} nonlinear plane wave which can be approximated by the harmonic ansatz
\be
\chi(x) = A [1+ h \cos \, \big ( \sqrt{\xi} \, Q A \;  n_\mu x^\mu +\alpha \big )]\label{4p50}
\ee
with an arbitrary unit vector $n_\mu$ and phase $\alpha$ \footnote{Clearly this is reminiscent of the toy model in [1] where it has been shown rigorously that the global minimum
is assumed for a single (harmonic, in this case) plane wave.}.
The constants $h\approx 0.58$ and $\sqrt{\xi}=\nu/AQ\approx 0.45$ are universal numbers independent of $\beta$.
The overall constant $A$ is not determined by the minimization condition but rather gets fixed when we adopt a normalization condition for the dimensionful
coordinates. Here it is natural to introduce 
\be
\bar{x}^\mu\equiv \sqrt{\frac{2\pi}{3}(2+h^2)} \, A\, \lp \, x^\mu \label{4p51}
\ee
When expressed in terms of these coordinates the line element related to (\ref{4p50}) reads
\be
ds^2 = \frac{2}{2+h^2}\Big [ 1+h \cos \Big ( \, \sqrt{\frac{\xi}{48 \pi(2+h^2)\beta}} \, \mp \, n_\mu \bar{x}^\mu+\alpha \Big) \Big]^2\, \delta_{\mu\nu} d\bar{x}^\mu d\bar{x}^\nu\label{4p52}
\ee
Averaging over  the harmonic oscillations we find a flat spacetime with $\langle d s^2 \rangle = \delta_{\mu\nu} d\bar{x}^\mu d\bar{x}^\nu$ again.
Note that the frequency of the oscillations increases for decreasing $\beta$. It approaches infinity in the limit of a pure Einstein-Hilbert action, $\beta\rightarrow 0$.
Note also that, since $h<1$, the conformal factor of (\ref{4p52}) has no zeros and the metric is everywhere non-degenerate.

\section{Impact of the cosmological constant: \\
numerical solutions}
To supplement the analysis of the previous section where we set $\Lambda$=0 we shall now allow for a nonzero cosmological constant and explore its impact 
on the stationary points. We restrict our attention to solutions depending on one cartesian coordinate only, $x^1\equiv x$, say. 
The equations (\ref{2p24}) and (\ref{2p28}) for $\chi$ and $\sigma$, respectively, are then most conveniently written in the form
\ba
&&\chi'' +\frac{u}{6}\chi^3 +\frac{72\beta}{\chi^4}\Big [\chi^2 \chi'''' - 3 \chi \, (\chi'')^2-4 \chi \, \chi' \, \chi''' + 6 \chi'' (\chi')^2\Big]=0\label{5p1}
\ea
and likewise 
\ba
&& \Big [\sigma''+(\sigma')^2 \Big ]\, e^{2\sigma}+\frac{u}{6}\,e^{4\sigma}+72\beta \Big [ \sigma''''-6 \, \sigma'' (\sigma')^2 \Big ]=0\label{5p2}
\ea
Here the prime denotes a derivative with respect to $x$. 

\subsection{Linear approximation}
Later on we shall use a numerical technique in order to find periodic solutions to these ordinary differential equations. Before embarking on that it is 
useful to analyze their linearization which describes small deviations from flat space, i.e. from $\chi=1$ or $\sigma=0$, respectively.
\begin{figure}[h]
\centering
\includegraphics[width=8cm]{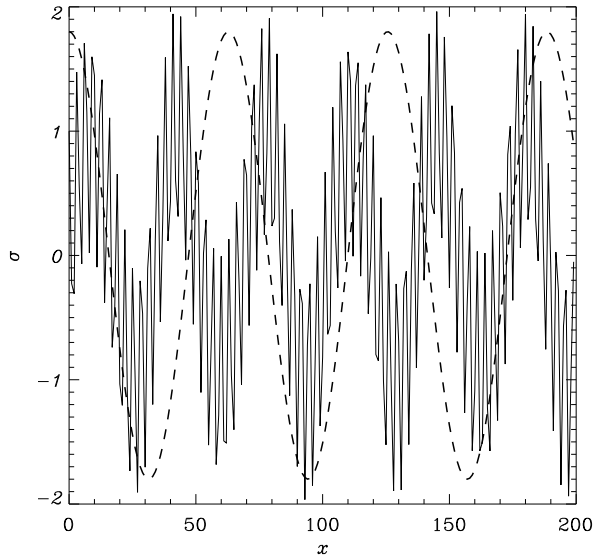}
\caption{
Analytical solution of eq.(\ref{5p3}) for $Q=2$ and $u=5 \cdot 10^{-2}$ (solid line). We observe a slow
periodic ``Hubble" evolution with frequency $\omega_{-}$, superimposed with rapid ``Planckian" oscillations with frequency $\omega_{+}$.
For comparison also the solution for $Q=10^{-1}$ and $u=0$ is shown (dashed line).  It represents a magnification of the single-scale
``Planckian" oscillations for vanishing cosmological constant; it could be regarded a model for Minkowski space at small distances.
\label{sol1}}  
\end{figure}

Starting from Eq.(\ref{5p2}) we expand $e^\sigma=1+\sigma+\dots$ and retain the derivative  
and non-derivative $O(\sigma^1)$ terms, as well as the  $O(\sigma^0)$ term. 
This leads to the linear equation
\be
\sigma'' +\frac{u}{6}(1+4\sigma)+72\beta \sigma''''=0\label{5p3}
\ee
Its most general solution reads
\be
\sigma(x) = -\frac{1}{4} +A_{+}\cos(x \, \omega_{+}) +A_{-}\cos(x\, \omega_{-})+B_{+}\sin(x\, \omega_{+}) +B_{-}\sin(x\, \omega_{-}) \label{5p4}
\ee
where $A_{\pm}$, $B_{\pm}$ are integration constants, and
\be
\omega_{\pm} = \frac{1}{\sqrt{6}}\sqrt{Q(3Q \pm \sqrt{9Q^2-24 u})} .
\ee
Recall also that $u\equiv \frac{16\pi}{3} G\Lambda $ and $Q^2\equiv1/(72\beta)>0$. 

The constants $\omega_{\pm}$ are real, and so $\sigma(x)$ is periodic if the cosmological constant is small enough, namely when $Q^2\geq 8 u/3$. Otherwise they have an
imaginary part which leads to an exponential behavior of the solution.

For an exactly vanishing cosmological constant $(u=0)$ we have $\omega_{+}=Q$ and $\omega_{-}=0$, whence $\sigma(x)$ is periodic with a single period $\propto 1/\beta$ 
determined by the $R^2$ term.

\noindent
When we switch on a small cosmological constant the ``large" frequency $\omega_{+} = Q$ is not affected much, but $\sigma(x)$ develops an additional periodicity 
with the ``small" frequency $\omega_{-}$; to leading order in $u$, we have $\omega_{-} \approx \sqrt{2u/3} \propto \sqrt{\Lambda}$.

Thus in general the solution (\ref{5p4}) displays two different scales on which it varies. If $Q^2\gg u$, $\omega_{-}\rightarrow 0$  and the only relevant frequency 
is the period of the by now familiar $R^2$-term.  For large $Q$-values the corresponding frequency is proportional
to $1/\sqrt{\beta}$, and since $Qx\propto \bar{x}/(\sqrt{\beta \lp})$ we see that dimensionful coordinate period $\Delta\bar x$ is of the order $\sqrt{\beta} \lp$, consistent with the
findings above. 

The second scale is set by the cosmological constant 
which can turn the oscillating solution (\ref{5p4}) into an exponentially one for $\Lambda$ large enough,  $Q^2<8u/3$ .

The overall behavior of $\sigma(x)$ is governed by the interplay of those two relevant length scales. Their interpretation is particularly clear when
$\beta= O(1)$ and $\Lambda\ll \mp^2$ since then $u/Q^2 \propto G\Lambda \beta\ll 1$. In this case the small period is of the order of the  
Planck length $\lp=\sqrt{G}$ while the large period is the ``Hubble" scale $\propto 1/\sqrt{\Lambda}$.
Hence eq.(\ref{5p4}) describes a slow cosmological evolution caused by $\Lambda$ superimposed with rapid oscillations on the Planck scale. An example of this situation 
is depicted in Fig.(\ref{sol1}). This picture  is confirmed by the numerical investigation discussed below.
\begin{figure}[h]
\centering
\includegraphics[width=6.5cm]{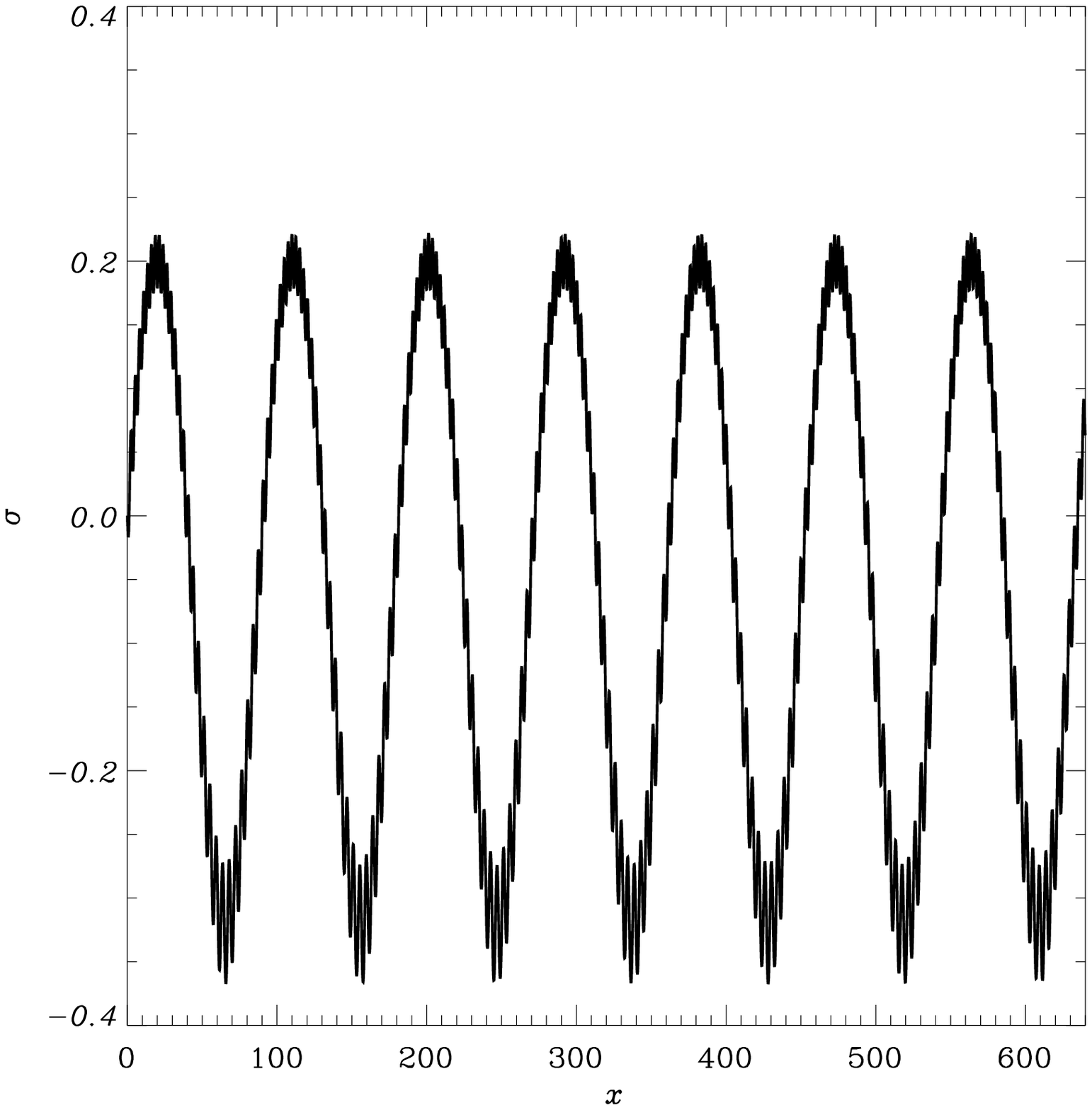}
\includegraphics[width=6.5cm]{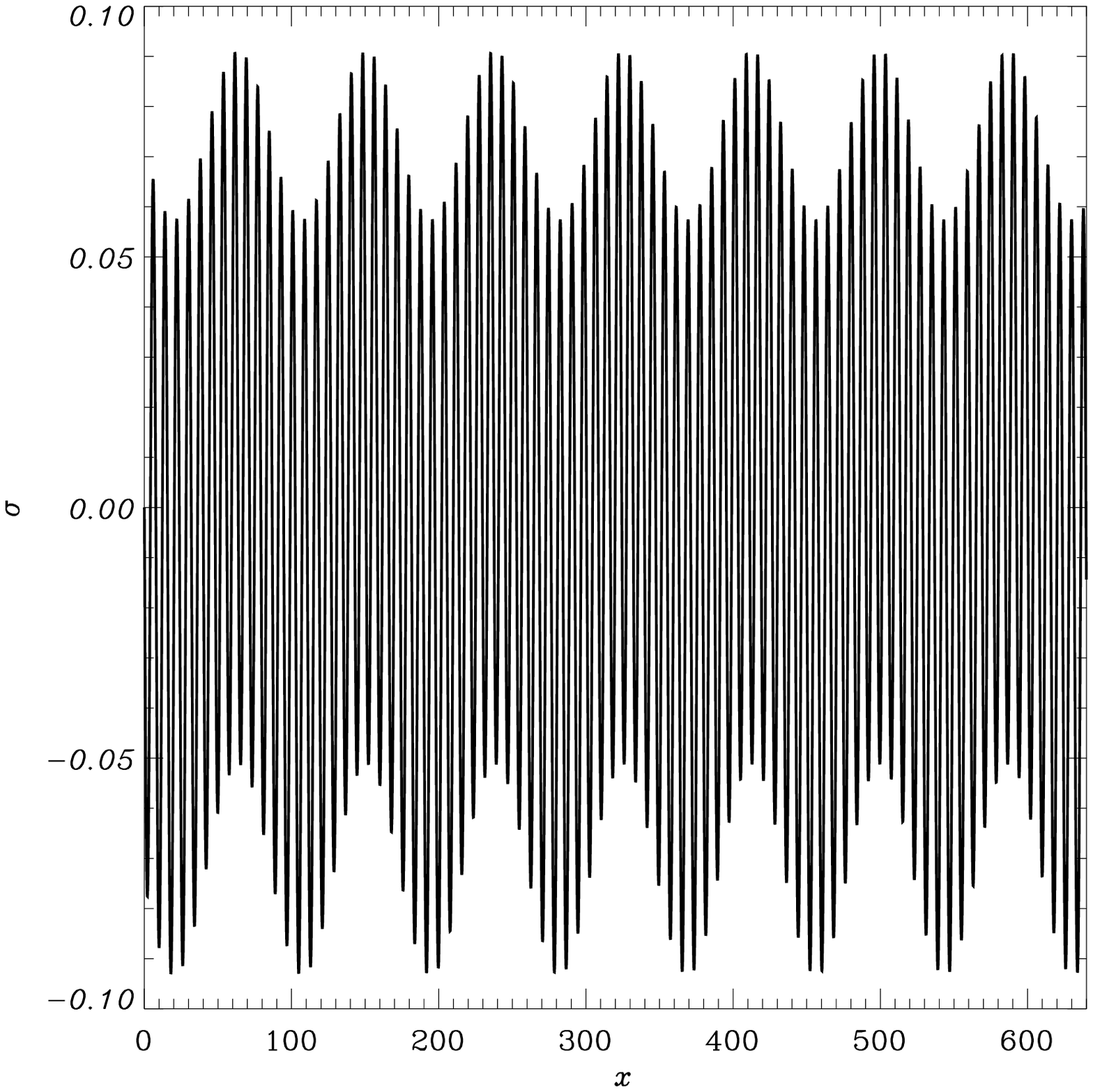}
\includegraphics[width=6.5cm]{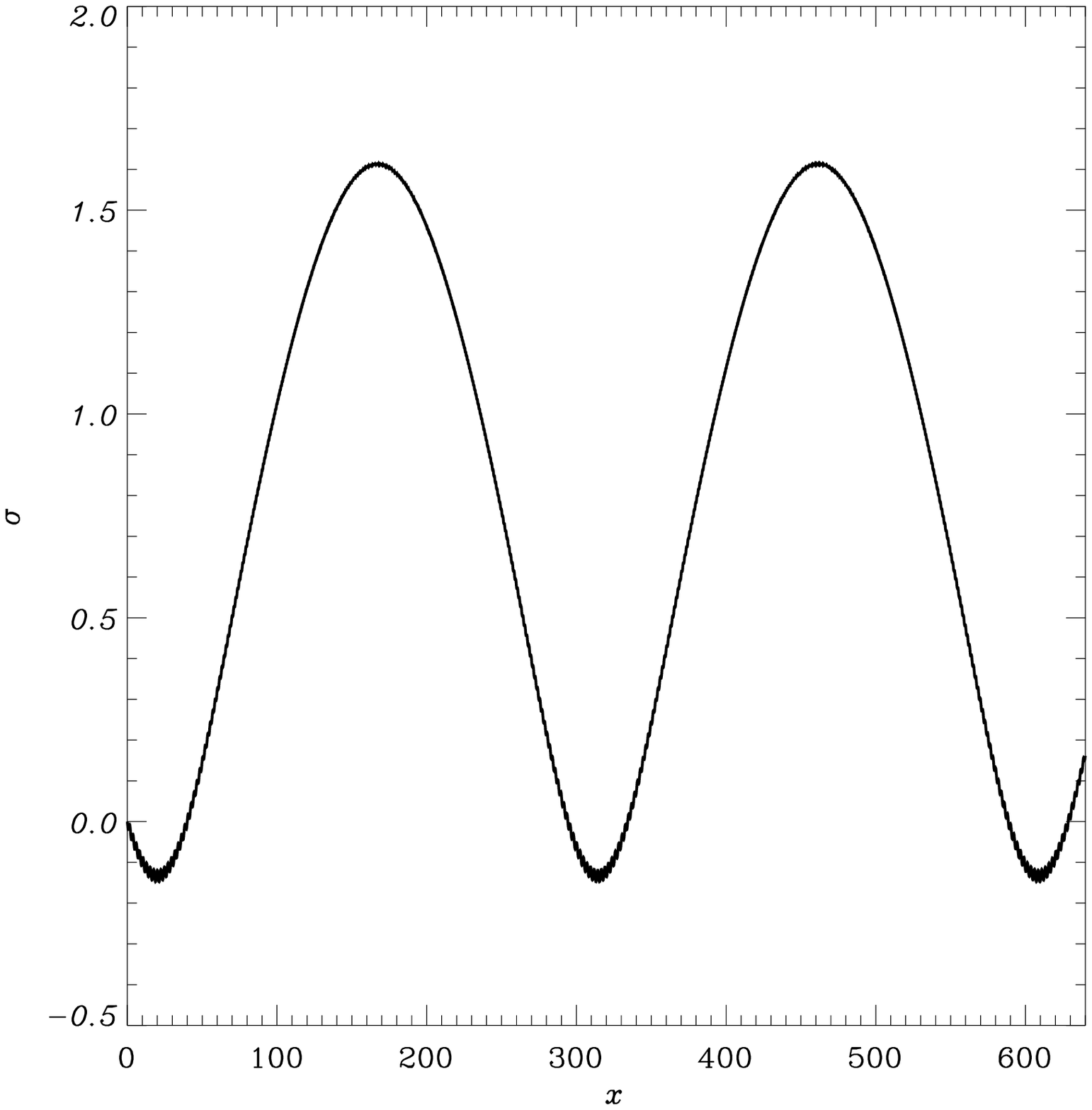}
\includegraphics[width=6.5cm]{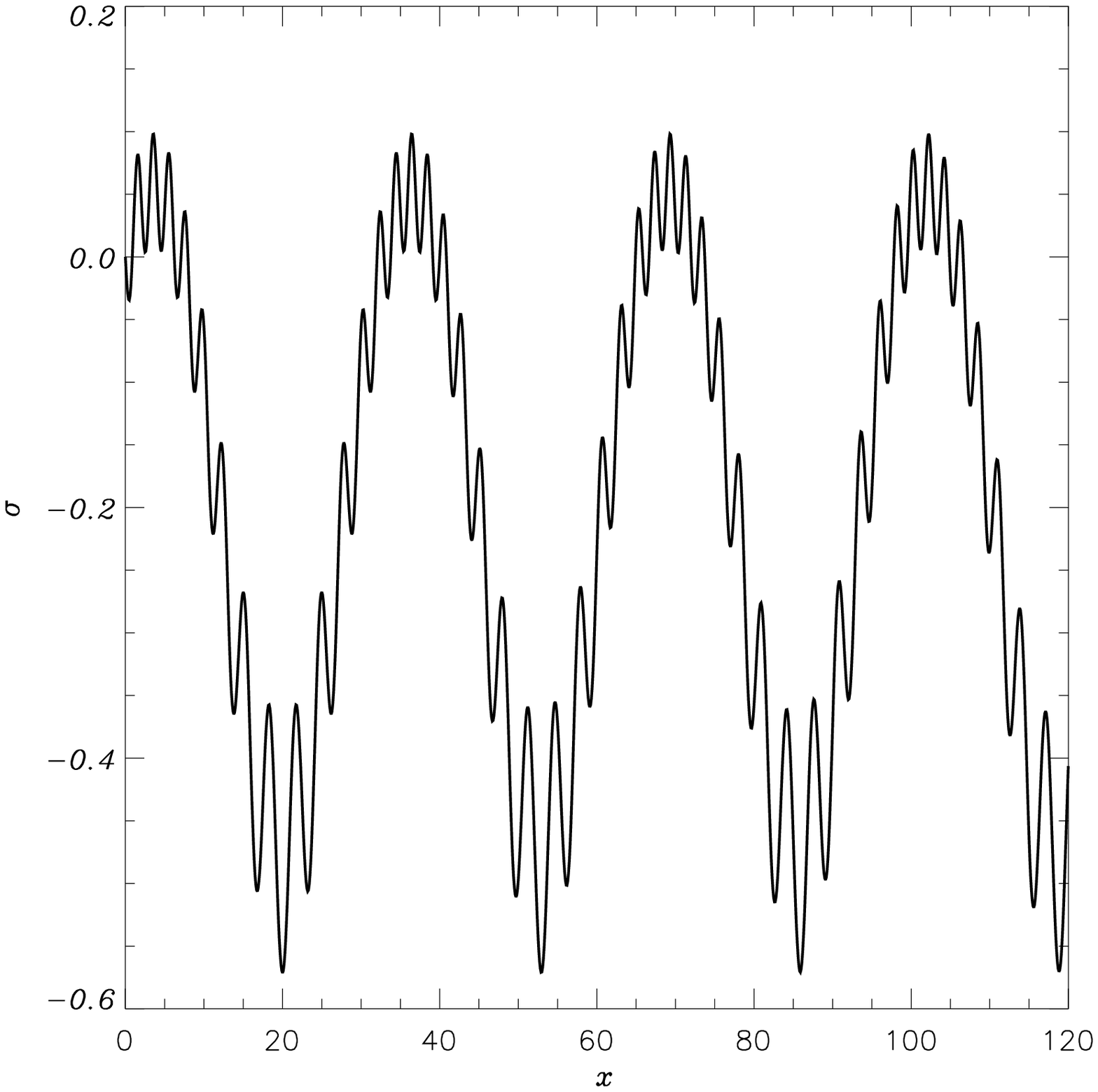}
\caption{Upper panel: numerical solution of eq.(\ref{5p2}) for $u=10^{-2}$ and $Q=2$ (left) and $Q=0.8$ (right).
Lower panel: numerical solution of eq.(\ref{5p2}) for $Q=3$ and $u=10^{-4}$ (left) and $u=10^{-1}$ (right). 
\label{Q3m2.eps}
}  
\end{figure}

If $Q^2<8u/3$ the general solution of (\ref{5p3}) is a linear combination involving growing and decaying exponentials which drive the solution outside
of the linear regime. To see what happens then, we ``switch off" the Planck
scale oscillations and set $\beta=0$ in (\ref{5p1}) of $\chi\equiv e^\sigma$. This results in  
\be
\chi''+\frac{u}{6}\chi^3=0\label{5p5}
\ee
which happens to be the one dimensional restriction of the Yamabe equation (\ref{3p2}) discussed earlier. (From the conceptional point of view this mathematical equivalence
is to some extent coincidential, however. Eq.(\ref{3p2}) had the interpretation of a Bogomolny-like equation for the minimum action configuration
{\it for the special parameter value} $\gamma\equiv 24 u \beta -1=0$ which requires $\beta\not=0$.
Only then the solution of the Bogomolny equation has the lowest possible action, $S=0$. In the case at hand, instead, we set $\beta=0$ so that (\ref{5p5}) relates to the 
simple Einstein-Hilbert action only.) 

Nevertheless, we know that (\ref{5p5}) has the periodic solution
\be
\chi(x) = \hat{\chi} \; {\rm sn} \, (\hat{\chi}\sqrt{u/12} \; x+\alpha; i)
\ee
with $u\propto G\Lambda = \lp^2 \Lambda$. Thus, introducing $\bar{x}\propto \lp x$ the first argument of the {\it sn} function is essentially $\sqrt{\Lambda \bar x}$. This nonlinear
oscillation, with a period $\Delta \bar x\propto 1/\sqrt{\Lambda}$ represents the generalization of the above ``slow", or ``cosmological" variation in the nonlinear regime.

This discussion suggests the following general structure of the stationary points in general (and also of the absolute minimum, hopefully). 
If for simplicity $\Lambda\ll \mp^2$, the conformal factor has a double periodicity; on small length scales it undergoes oscillations
(with period $\Delta\bar x \propto \sqrt{\beta }\lp$) which are due to the higher derivative term in the action. These oscillations are superimposed on another type of oscillations
of a much larger period set by the cosmological constant: $\Delta \bar x\propto 1/\sqrt{\Lambda}$.
This feature, too, is confirmed by the numerical investigations to which we turn next.

\subsection{Numerical solutions}
We are in particular interested in numerically determining the periodic solutions of Eq.(\ref{5p2}). In order to achieve this goal it is convenient 
to follow the method presented in \cite{pelet} and to consider 
(\ref{5p2}) as  the {boundary value problem} 
defined by the conditions
\begin{subequations}
\label{bc}
\ba
&&\sigma(x_1) = \sigma''(x_1) = 0  \\[2mm]
&&\sigma(x_2)=\sigma''(x_2)=0
\ea
\end{subequations}
being $x_1$ and $x_2$ the initial and final limits  of the integration interval, respectively. 
A problem of this type can be conveniently solved by means of the {shooting method} embedded in a globally convergent 
Newton-Rawson algorithm. In other words, the conditions at the outer integration extremum $x_2$ are mapped in a functional dependence 
of the initial conditions for $\sigma'=\sigma'(x=x_1)$ and $\sigma'''=\sigma'''(x=x_1)$ in order to satisfy (\ref{bc}) at $x=x_2$. 
Once convergence is achieved, it is possible to adiabatically explore the parameter space spanned by the two variables $Q$ and $u$.

The results are summarized in Fig.(\ref{Q3m2.eps}). In particular in the two upper panels the effect of decreasing $Q$ (which means increasing $\beta$) while
keeping $u$ fixed is shown, so that the frequency of the small scale $R^2$-induced fluctuations increases  as $Q$ decreases (right upper panel).
On the other hand, if we instead keep $Q$ fixed and change $u$ (lower panels), the large scale non-linear periodicity emerges as $u$ is decreased.

\begin{figure}[t]
\centering
\includegraphics[width=10cm]{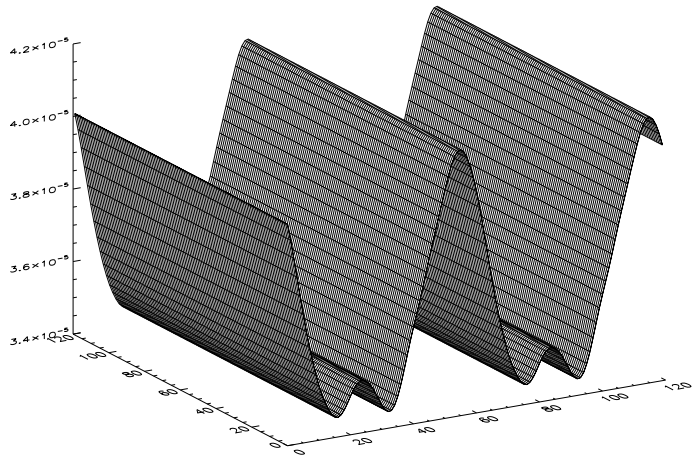}
\caption{Minimum action configuration of the lattice regulated action (\ref{preno}) obtained for $\beta=27$ and $u=0$.
\label{f1}}  
\end{figure}

\begin{figure}[t]
\centering
\includegraphics[width=10cm]{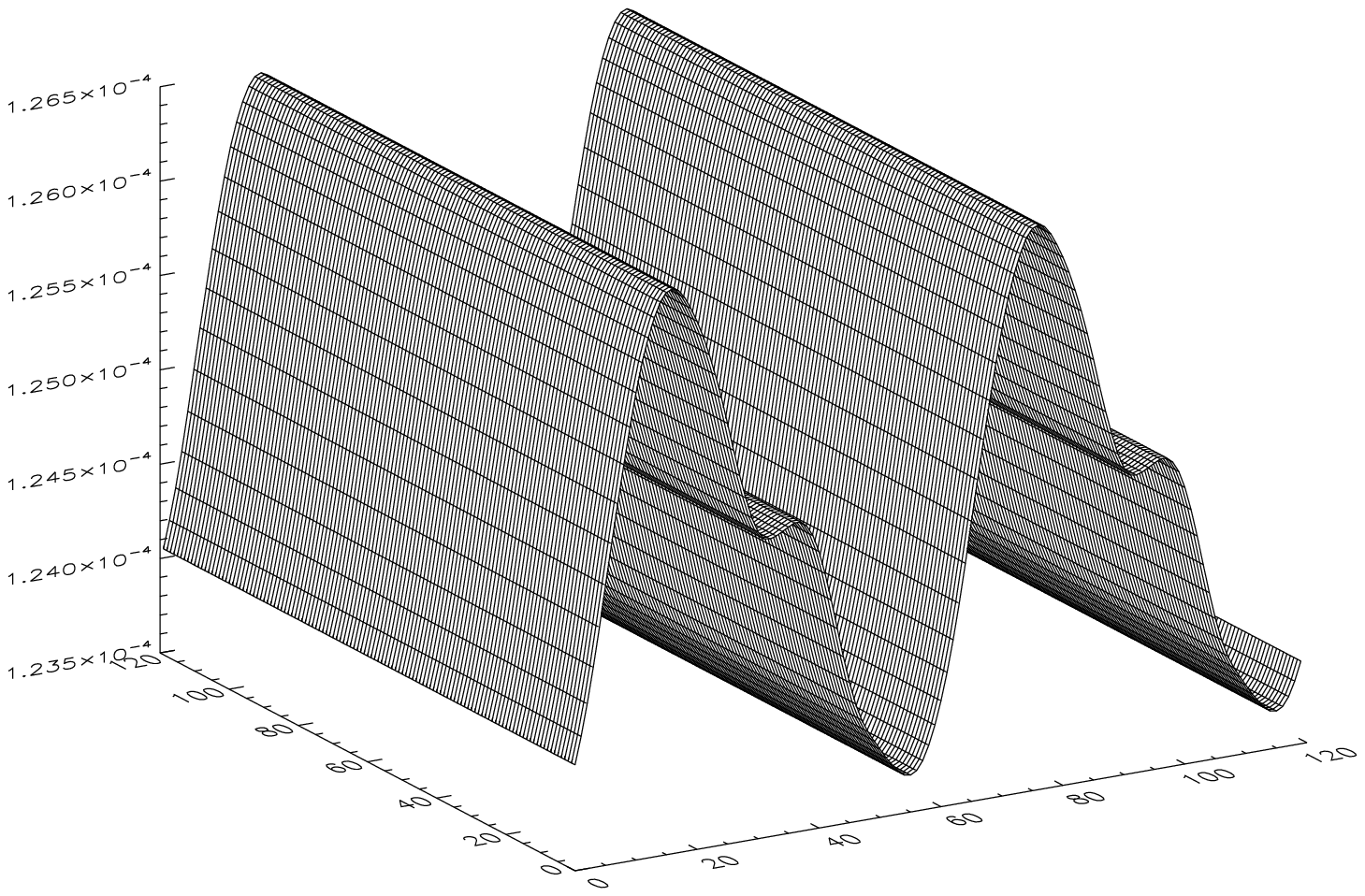}
\caption{
Minimum action configuration of the lattice regulated action (\ref{preno}) obtained for $\beta=55$ and $u=0$.
\label{f2}
}  
\end{figure}
\begin{figure}[t]
\centering
\includegraphics[width=10cm]{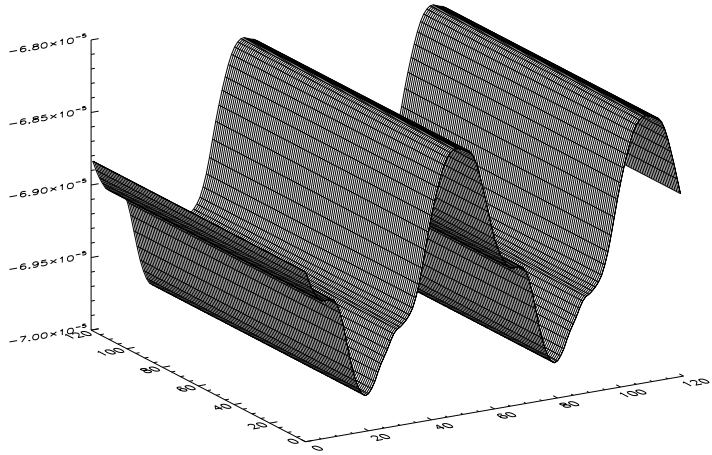}
\caption{
Minimum of the lattice regulated action (\ref{preno}) obtained for $\beta=41$ and $u=0.055$.
\label{f3}
}  
\end{figure}

\subsection{Lattice regulated model}
In more than one dimension the task of determining the general solution of  (\ref{2p24})  or (\ref{2p28}) is extremely hard due to their elliptic non-linear 
structure. A possible strategy is to directly minimize the lattice regulated version of the functional  (\ref{2p27}) as a multivariate function of the field at each lattice site:
\ba
&&S[\sigma(x)]=\sum_x   \Big\{ \frac{u}{4!} \;  e^{4\sigma(x)} \nonumber\\ 
&&+\sum_\mu \frac{1}{2} \Big [ e^{2\sigma(x)} (\sigma(x+e_\mu) +\sigma(x-e_\mu) -2 \sigma(x) +  (\sigma(x+e_\mu) -\sigma(x))^2)\nonumber\\
&&+\sum_\nu 36 \beta (\sigma(x+e_\mu) +\sigma(x-e_\mu) -2 \sigma(x))   (\sigma(x+e_\nu) +\sigma(x-e_\nu) -2 \sigma(x)) \nonumber\\
&&+(\sigma(x+e_\mu) +\sigma(x-e_\mu) -2 \sigma(x)) (\sigma(x+e_\nu) -\sigma(x))\nonumber\\
&&+(\sigma(x+e_\nu) +\sigma(x-e_\nu) -2 \sigma(x)) (\sigma(x+e_\mu) -\sigma(x))\nonumber\\
&&+(\sigma(x+e_\mu) -\sigma(x)) (\sigma(x+e_\nu) -\sigma(x))^2\Big ]\Big \}\label{preno} .
\ea
In (\ref{preno}) the variable $\sigma$ is a dimensionless field so that the lattice cutoff is $a\equiv 1$, and  ${e_\mu}^\nu = \delta_{\mu}^{\nu}$.
For computational reasons we work in $d=2$ dimensions and we consider a two-dimensional lattice. 

The size of the lattices used
ranges from $60\times 60$ mesh points up to $120 \times 120$ mesh points, in order to 
check the numerical stability of the results. For actual calculations the Fletcher-Reeves conjugate gradient algorithm, 
which implements a succession of line minimizations, turned out to be particularly convenient. After an initial search generated by a uniform random field distribution,  
the direction is chosen using the gradient of the  action. The line minimization is thus carried out iteratively in that direction \cite{gnu} and
the algorithm works quite well when the period of the $R^2$ term is the dominant one.  

In Fig.(\ref{f1}) and Fig.(\ref{f2}) the field configuration corresponding to the global minimum of the action is displayed
for $u=0$, $\beta=27$ and $\beta=55$ respectively, corresponding to $\Lambda =0$ .
It is reassuring to notice that it is an essentially one-dimensional object, having a nontrivial dependence on one coordinate only.
This proves, at least for $d=2$, that the global minimum is a single non-linear 
plane wave, in complete agreement with our previous variational calculation.  

Fig.(\ref{f3}) is for $\Lambda\not =0$ and shows instead the result of the minimization obtained for   $\beta=41$ and $u=0.005$. 
Here, too, the global minimum action configuration consists of a single non-linear 
plane wave with a double-periodicity. Further exploration of the parameter space shows that generically the 
possible ``zoo" of stationary configurations has this very special feature.


\section{Conclusions}
In this paper we explored the possibility that the conformal factor instability of the Einstein-Hilbert action is cured by an additional higher derivative
invariant. As a concrete realization of this idea we considered $\Lambda+R+\beta R^2$ gravity in 4 dimensions, for technical reasons restricted to the
purely conformal sector. Using various techniques we found that the field configurations which correspond to the 
global 
minimum of the relevant action functional is a family of non-linear plane waves
which are labeled by a unit vector $n^\mu$ and a phase angle $\alpha$. While the field is constant on the hyperplanes perpendicular to $n^\mu$, 
it has a nontrivial modulation along the direction of $n^\mu$.
Approximately, when the cosmological constant is small, this modulation can be thought of as the superposition of two harmonic waves with two
frequencies determined by the cosmological constant and the coefficient of $R^2$, respectively. If their scales are well separated, we are led to the picture
of Planckian ripples due to the $R^2$ term which are superimposed on a much smoother solution to the simple $\Lambda+R$ theory.

This result will be a key ingredient in the analysis of the ground state of the quantized theory. Following the approach of \cite{larewe} the logical next step
will consist in a computation of the pertinent effective action $\Gamma$ by means of a saddle point expansion about the global minimum which we found.
Since this minimum is degenerate with respect to $n^\mu$ and $\alpha$, this computation will involve an integration over the moduli space $S^3\times S^1$
already at the leading order of the semiclassical expansion, thus restoring full $O(4)$ invariance at the level of $\Gamma$. Since the saddle point is known only 
numerically, this is a rather difficult step; we shall came back to it elsewhere.

The effective action functional $\Gamma$ will in particular contain information about the theory's ground state and small fluctuations around it. 
If the semiclassical expansion is valid we may expect this ground state to be symmetry breaking, i.e. modulated, picking a specific point in the vacuum manifold $S^3\times S^1$.

There is an important  analogy between the underlying stabilization mechanism of the conformal factor instability
and known examples of modulated phases in Solid State Physics.
In particular, in either case spatially nonconstant modes condense as the result of the competing (negative) nearest-neighbor kinetic term and the (positive)
next-to-nearest-neighbor term which, in gravity, stems from the four derivatives in $R^2$.

It would be nice to discuss possible experimental or observational signatures of this phenomenon, along the lines
suggested in \cite{per}, and we hope to address this issue in a following work.

\end{document}